\documentstyle[psfig,twocolumn,aps]{revtex}
\newcommand{\livo}{Li$_2$VOSiO$_4$}
\newcommand{\ligo}{Li$_2$VOGeO$_4$}
\begin{document}
\draft
\wideabs{ \title{ 
Realization of a large $J_2$ quasi-2D spin-half Heisenberg system: \livo}
\author{H.\ Rosner\cite{dre}$^{,1}$, R.R.P.\ Singh$^{1}$, W.H. Zheng$^2$,
J. Oitmaa$^2$, S.-L.\ Drechsler$^3$, and W.E.\ Pickett$^1$}
\address{$^1$Department of Physics, University of California, Davis, CA
95616, USA}
%
%
\address{$^2$School of Physics, University of New South Wales, Sydney NSW 2052, Australia}
%
%
\address{$^3$Institut f\"ur Festk\"orper- und Werkstofforschung Dresden e.V.,
Postfach 270116, D-01171 Dresden, Germany}
\date{\today}
\maketitle
\begin{abstract}
Exchange couplings are calculated for \livo\ using LDA.
While the sum of in-plane couplings $J_1 + J_2 = 9.5 \pm 1.5$K and
the inter-plane coupling $J_\perp \sim 0.2$--$0.3$K agree with recent
experimental data, the ratio $J_2/J_1 \sim 12$ exceeds the reported
value by an order of magnitude.  Using geometrical considerations,
high temperature expansions and perturbative mean field 
theory, we show that the
LDA derived exchange constants lead to a remarkably accurate
description of the properties of these materials
including specific heat, susceptibility, Ne\'el temperature and NMR
spectra.
\end{abstract}
}
\narrowtext

In many recently discovered magnetic materials the determination of
exchange constants, without input from electronic structure
calculations, has proven very difficult and has often led to wildly
incorrect parameter values.  The interplay of geometry and quantum
chemistry has yielded many surprises which could not have been
anticipated without a full calculation.  Examples are the recently
discovered vanadates CaV$_4$O$_9$\cite{pickett} and
CaV$_3$O$_7$\cite{korotin}.  In all these cases the dominant exchange
interactions were resolved and a good understanding of the material
properties obtained only after analyses of
electronic structure calculations were carried out.


Frustrated square-lattice spin-half Heisenberg antiferromagnets
with nearest neighbor exchange $J_1$ and second neighbor 
(diagonal) exchange
$J_2$ have received considerable attention recently. The properties
of the model with $J_2$=$0$ (or $J_1$=$0$) are well understood at zero
and finite temperature \cite{CHN}. The large $J_2$ limit of the model is a
classic example of quantum order by disorder\cite{Shender,Chandra}, where at the
classical level the two sublattices order antiferromagnetically but
remain free to rotate with respect to each other. This degeneracy is
lifted by quantum fluctuations leading to collinear magnetic order in a
columnar pattern. At intermediate $J_2/J_1$ there is strong evidence
for a spin-gap phase, though the nature of this phase is not fully
resolved yet \cite{Sushkov}.

While there has been tremendous theoretical interest in these models,
there were no known experimental realizations for intermediate to
large $J_2/J_1$, until the investigation of \livo\ by Melzi {\it et
al.}\cite{melzi00,melzi01} Studying the splitting patterns of the
$^7$Li NMR spectra, these authors presented strong evidence for
columnar order\cite{melzi00}. Combining several experiments they
derive\cite{melzi01} exchange couplings (with $J_2/J_1 \sim 1.1$) well
into the region where model calculations find columnar order.

However, several puzzling pieces in that excellent and detailed
study remain: (i) The ratio of exchange
constants was not well determined from the susceptibility and specific
heat data; we will present electronic structure and many-body
calculations to show that their estimate $J_2/J_1\approx
1$\cite{melzi01}, is off by an order of magnitude. (ii) The estimated $T=0$ moment
was anomalously small for a system well inside the columnar ordered
phase. Taking into account the antiferromagnetic inter-plane coupling,
we propose that the NMR derived moment is small due to a cancellation
of hyperfine fields from neighboring planes.
(iii) The order parameter exponent $\beta$ at the transition was
estimated  to be $\beta\approx 0.25$, which is
intermediate between 2D Ising and typical 3D exponents. We will show
that the inter-plane exchange constants differ from the largest ones by
less than two orders of magnitude. Thus a strong crossover between 2D and 3D
behavior could be expected. (iv) The Ne\'el temperature was nearly
field independent up to a field of $9T$. We will argue that our
increased estimate of $J_2$ leads to a larger saturation field
and that combined with non-monotonic dependence of Ne\'el temperature
on field implies that the experimental results are not anomalous.

Our study of the material \livo\ consists of
a two band tight-binding model fit to the LDA band structure,
which is then mapped onto a Heisenberg model with in-plane 
($J_1$ and $J_2$) and 
inter-plane ($J_\perp$) exchange constants. Furthermore,
we develop high temperature series expansions and
perturbative mean-field theory for the
uniform susceptibility and specific heat of the $J_1-J_2$ model.
These allow us to make quantitative comparisons with the experiments.

\begin{figure}[bt]
\begin{center}
\begin{minipage}{7.5cm}
\psfig{figure=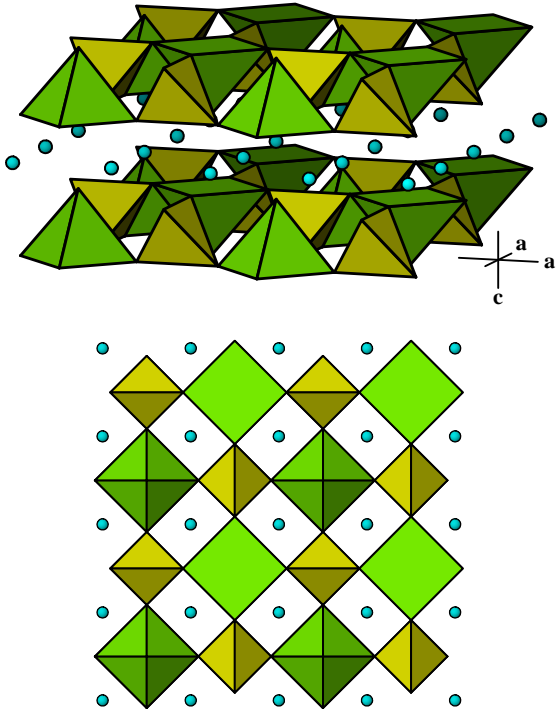,width=7.5cm}
\end{minipage}
\end{center}
\vspace{3mm}
\caption{
Perspective view (upper panel) of the crystal structure of \livo\, and projection
along [001] (lower panel). The VO$_5$ pyramids (large diamonds) 
share the corners of the
basal planes with SiO$_4$ tetrahedra (small diamonds). 
The Li$^+$ ions are indicated by
circles.
}
\label{struct}
\end{figure}

\livo\ crystallizes in the tetragonal system, space group $P4/nmm$,
containing two formula units per cell with $a = 6.3682$ \AA\ and $c =
4.449$ \AA.\cite{satto98} The crystal structure of \livo\ is shown in
Fig.~\ref{struct}. The magnetically active network of spin half
V$^{4+}$ ions is built up by [VOSiO$_4$]$^{2-}$ layers of VO$_5$
square pyramids sharing corners with SiO$_4$ tetrahedra, intercalated
with Li ions. The structure of the V$^{4+}$ square network suggests,
that both the nearest neighbor (NN) and the next nearest neighbor
(NNN) in-plane coupling should be significant, although it is at
best difficult to decide from general considerations which one is
dominant. NN coupling is favored by the existence of two exchange
channels and shorter distance, NNN coupling profits from the
'straight' connection between pyramids pointing in the same direction.

In order to obtain a realistic and reliable hopping part of a tight-%
binding Hamiltonian, band structure calculations were performed using
the full-potential nonorthogonal local-orbital minimum-basis scheme
\cite{koepernik99} within the local density approximation (LDA). In
the scalar relativistic calculations we used the exchange and
correlation potential of Perdew and Zunger\cite{perdew81}.
V($3s$,$3p$,$4s$, $4p$, 3$d$), O(2$s$, 2$p$, 3$d$), Li(2$s$, 2$p$) and
Si(3$s$, 3$p$, 3$d$) states, respectively, were chosen as the
basis set. All lower lying states were treated as core states. The
inclusion of V (3$s$,3$p$) states in the valence states was necessary
to account for non-negligible core-core overlaps. The O and Si 3$d$ as
well as the Li 2$p$ states were taken into account to increase the
completeness of the basis set. The spatial extension of the basis
orbitals, controlled by a confining potential \cite{eschrig89}
$(r/r_0)^4$, was optimized to minimize the total energy.

\begin{figure}[bt]
\begin{center}
\begin{minipage}{7.5cm}
\psfig{figure=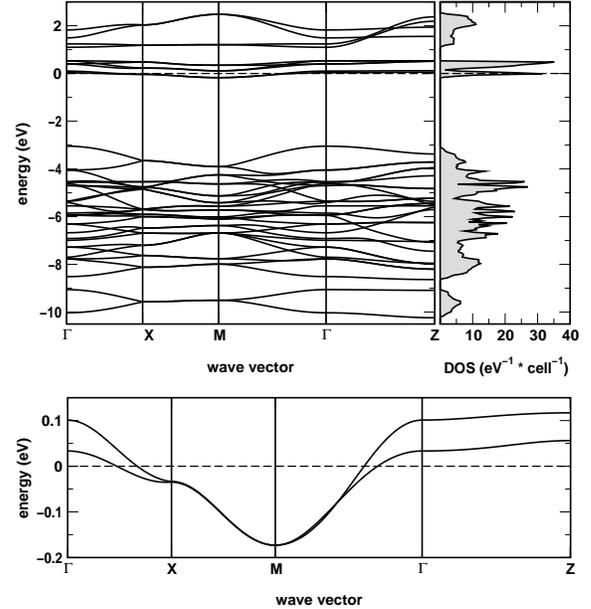,width=7.5cm}
\end{minipage}
\end{center}
\vspace{3mm}
\caption{ Band structure and total density of states for \livo\ (upper
panel) and the zoomed bands closest to the Fermi level (lower
panel). The Fermi level is at zero energy. The notation of the
symmetry points is as follows: X = (100), M = (110), Z = (001)
}
\label{band}
\end{figure}

The results of the paramagnetic calculation (see Fig.~\ref{band}) show
a valence band complex of about 10 eV width with two bands crossing
the Fermi level. These two bands, due to the two V per cell, are well
separated by a gap of about 3 eV from the rest of the valence band
complex and show mainly V 3$d_{xy}$ and minor O(2) 2$p_{x,y}$
character (oxygens of the basal plane of the VO$_5$ pyramid) in the
analysis of the corresponding orbital-resolved partial densities of
states (not shown). The valence bands below the gap and above the
Fermi level have almost pure oxygen and vanadium character,
respectively. The contribution of Li and Si states is negligible in
the energy region shown.

The relatively narrow bands at the Fermi level (see Fig.~\ref{band},
lower panel) are half-filled.  Therefore, strong correlation effects
can be expected which explain the experimentally observed insulating
ground state. Because the low-lying magnetic excitations involve only
those orbitals with unpaired spins corresponding to the half-filled
bands, we restrict ourselves to a two band tight-binding analysis
and the discussion of these half-filled bands.

The dispersion of these bands (see Fig.~\ref{band}, lower panel) has
been analyzed in terms of NN transfer $t_1$ and NNN transfer $t_2$
within the [001] plane (see Fig.~\ref{struct} lower panel) and NN
hopping $t_\perp$ between neighboring planes.

Then, the corresponding dispersion relation of the related 2$\times$2
problem takes the form
\begin{eqnarray}
E(\vec{k})&=&\varepsilon_0 + 2t_2 [ \cos(x)+\cos(y) ] 
\nonumber \\
&&\pm 4t_1\cos(x/2)\cos(y/2)
+ 2t_\perp\cos(z)\, ,
\end{eqnarray}
where $x=k_za$, $y=k_yb$, $z=k_zc$.  

The assignment of the parameters has been achieved by two numerically
independent procedures: By straightforward least square fitting of the
two bands in all directions and by using the energy eigenvalues at
different selected high symmetry points.  The results are shown in
Table~\ref{table1}. The errors can be estimated about 5\% for the
in-plane transfers and 15\% for the inter-plane term from the
differences of both mentioned above fitting procedures. These small
differences can be ascribed to the influence of higher neighbors. The
very good agreement of the tight binding fit with the LDA bands
justifies {\it a posteriori} the restriction to NN and NNN couplings only.
\vbox{
\begin{table}
{\begin{tabular}{|c|c|c||c|c|c|c|}
$t_{1}$ (meV) &
$t_{2}$ (meV) &
$t_{\perp}$ (meV) &
$ U $ (eV)&
$J_{1}$ (K) &
$J_{2}$ (K)&
$J_{\perp}$(K)\\
\hline 
8.5 &
29.1 &
-4.8 &
4&
0.83&
9.81&
0.27\\
&
&
&
5&
0.67&
7.85&
0.22\\
\end{tabular}\par}\vspace{.5cm}
\caption{ \label{table1}
Transfer integrals of the two-band tight-binding model and the
corresponding exchange couplings for different values of the Hubbard
$U$.
}
\end{table}
}
The resulting transfer integrals enable us to estimate the relevant
exchange couplings, crucial for the derivation
and examination of magnetic model Hamiltonians of the spin-1/2
Heisenberg type:
\begin{equation}
H_{spin}={\sum_{ij}}J_{ij}\vec{S_i}\cdot \vec{S_j} \, .
\end{equation}
In general, the total exchange $J$ can be divided into an
antiferromagnetic and a ferromagnetic contribution $J$ = $J^{AFM} +
J^{FM}$.  In the strongly correlated limit, valid for typical
vanadates, the former can be calculated in terms of the one-band
extended Hubbard model $J^{AFM}_{i}$ = $4t^2_{i}/(U-V_{i})$. The index
$i$ corresponds to NN and NNN, $U$ is the on-site Coulomb repulsion
and $V_{i}$ is the inter-site Coulomb interaction. Considering the
fact that the VO$_5$ pyramids are not directly connected, but via
SiO$_4$ tetrahedra, ferromagnetic contributions $J^{FM}$ are expected
to be small. For the same reason, the inter-site Coulomb interactions
V$_i$ should be small compared with the on-site repulsion $U$.  From
LDA-DMFT(QMC) studies\cite{held01} and by fitting spectroscopic data
to model calculations\cite{mizokawa93}, $U \sim 4$--$5$ eV is estimated
for typical vanadates. Therefore, we adopt $U$=$4$ eV and $U$=$5$ eV as
representative values to estimate the exchange constants and their
sensitivity to $U$.  The calculated values for the exchange integrals
are given in Table~\ref{table1}.

Comparing our calculated exchange couplings with the experimental
findings\cite{melzi01}, we find excellent agreement for the sum $J_1 +
J_2 = 9.5 \pm 1.5$ K \cite{remark1} of the in-plane couplings,
reported from susceptibility data\cite{melzi01} to be $J_1 + J_2 = 8.2
\pm 1$ K. In contrast, we find a ratio $J_2/J_1 \sim 12$ which exceeds
the experimentally derived ratio in Ref.~\onlinecite{melzi01} $J_2/J_1
\sim 1.1 \pm 0.1$ by an order of magnitude.

\begin{figure}[bt]
\begin{center}
\begin{minipage}{6.5cm}
\psfig{figure=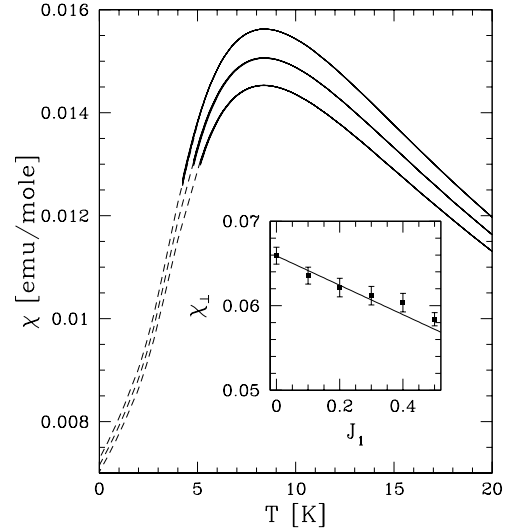,width=6.5cm}
\end{minipage}
\end{center}
\vspace{3mm}
\caption{
Susceptibility ($\chi$, with largest $\chi$ for $J_1=0$) for 
$J_2=9K$, $g=2$ and $J_1/J_2=0, 0.1, 0.2$.
The low temperature data is obtained
from QMC combined with Eq.~3, while the high temperature data comes from
HTE. The inset shows Ising series expansion calculations and Eq.~3 for 
$T=0, J_2=1$.
}
\label{chi-fig}
\end{figure}

In order to understand the experiments better,
we turn to high temperature expansions for the 
susceptibility and specific heat of the Heisenberg models.
Using series expansions ($T$=0) \cite{Ising}, 
non-linear sigma model theory \cite{CHN} (very
low-$T$), quantum Monte Carlo (QMC) simulations (low-$T$) \cite{QMC} and high
temperature expansions(HTE) (high-T) \cite{HTE}, the susceptibility of the
nearest-neighbor model ($J_1=0$) is known
accurately for all $T$.  Letting $J_2=1$ and
treating $J_1$ perturbatively, analogous to
chain mean-field theories,\cite{SIP} leads to the expression
\begin{equation}
\chi(J_1,T)=\chi_0(T)[1 - 4 J_1 \chi_0(T) ]
\end{equation}
where $\chi_0$ is the susceptibility for the Heisenberg model ($J_1=0$).
As shown in the inset of Fig.~\ref{chi-fig}, at $T=0$
for small $J_1/J_2$, this expression 
compares very well with the susceptibility calculated
from Ising series expansions.\cite{Ising} 
Fig.~\ref{chi-fig} also shows that applying Eq.~3 to the finite-$T$
QMC data for $\chi_0$ leads to susceptibility values which
join smoothly with the high-temperature expansion results.
Thus, we have accurate calculations for the susceptibility
of the model with small $J_1/J_2$ at all $T$. 

Rather than find a fit for the exchange constants, in
Fig.~\ref{chi-fig} we show the susceptibility with $g=2$; $J_2=9K$;
and $J_1/J_2=0, 0.1$ and $0.2$.  The results are close to experimental
values \cite{melzi00}.  We note that the agreement will be improved by
going to the lower limit of the calculated exchange constants and
slightly larger $g$-values.

The specific heat data was the primary source for the $J_1/J_2\approx
1$ conclusion by Melzi et al. \cite{melzi01}. They found that the peak
value of the specific heat in \livo\ was $0.436(4)R$ at $T^m=3.5(1)$.
We find that for the pure Heisenberg model the specific heat peaks at
$T^m=0.60(4)J$ with a peak value of $0.455(10)R$, in agreement
with Ref.~\onlinecite{makivic91}.  With small $J_1/J_2$ the peak
shifts to lower temperature and the specific heat becomes flatter. The
fact that the values for the pure Heisenberg model are close to the
experiments strongly favors a small $J_1$.

\begin{figure}[bt]
\begin{center}
\begin{minipage}{6.5cm}
\psfig{figure=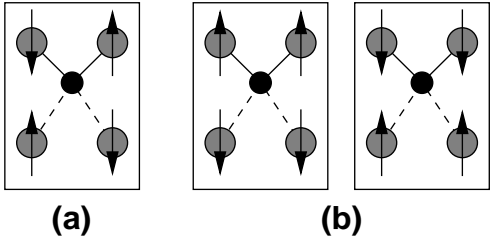,width=6.5cm,angle=-90}
\end{minipage}
\end{center}
\caption{ Sketch of the different magnetic environments for the $^7$Li
NMR. The Li and V sites are represented by black and gray circles,
respectively.  The arrows indicate the direction of the V spin. Full
lines symbolize the stronger interaction with the upper plane (NN),
dashed lines the weaker interaction with the lower plane (NNN). The
different environments cause (a) no NMR shift due to complete in-plane
moment-cancellation; (b) up or down shift with partial inter-plane
moment-compensation.
}
\label{cancel}
\end{figure}

One of the most puzzling aspects of the experimental
results\cite{melzi01} is the small moment of 0.24 $\mu_B$ at $T=0$,
obtained from the NMR split patterns. In
contrast, the moment of the square-lattice Heisenberg model is well
known to be $\approx 0.6 \mu_B$.\cite{Ising} 
Taking into account the considerable
antiferromagnetic inter-plane coupling $J_\perp$ resulting from our
calculation, a part of the discrepancy could be understood: The Li
nuclei sit off-center between two planes, which results in a partial
cancellation of the hyperfine fields from antiferromagnetically
ordered NN and NNN V sites in neighboring planes (see
Fig.~\ref{cancel}). This partial cancellation does not change the
arguments of Melzi {\it et al.} for the pattern of line-splitting
(including intensities) and its relation to columnar order because the
ordering pattern inside the planes remains the same. However, it leads
to a reduction in the effective hyperfine coupling and hence to an
enhancement of magnetic moment derived from the line shift.  Taking
into account the calculated two center overlap integrals for Li and NN
and NNN V 3d orbitals, respectively, (see Fig.~\ref{cancel}) a crude estimate from
Slater-Koster integrals suggests that the NMR split would be reduced
by an additional factor of about 2. This results in a moment of about
0.5 $\mu_B$ much closer to the value expected for the 2D Heisenberg
model.

We now turn to the inter-plane couplings and the measurements of the
Ne\'el temperature, $T_N$. Applying the expression $T_N\approx
0.36J_\perp \xi^2(T_N)$ \cite{CHN} ($\xi$ is the in-plane correlation
length), to our calculated exchange
constants, leads to the estimate $T_N\approx 3.6\pm 0.4$ K, which is
remarkably close to the experimental value of $2.8$ K. Furthermore, the
saturation field for our calculated exchange constants is about 30 T,
which is much bigger than the 9 T field applied by Melzi {\it et al.}
The Ne\'el temperature should go to zero at the saturation field.
However, we note that due to suppression of spin fluctuation
the Ne\'el temperature 
can increase
slightly with field, as happens in the purely 2D model. Thus, the
experimental result of very weak field dependence of the Ne\'el
temperature up to 9 Tesla is consistent with our expectations. The
appreciable but still small 3D couplings should also give rise to 3D
critical behavior at the finite temperature transition with strong
crossover effects. These results on the field dependence of the Ne\'el
temperature and the critical behavior at the transition in weakly
coupled Heisenberg systems deserve further theoretical attention.

To summarize, we have used LDA to calculate exchange constants for the
material \livo\ and developed numerical studies for the Heisenberg
model to show remarkable consistency with many experimental
properties.  Electronic structure calculations on
the closely related material \ligo ~will be presented in
a forthcoming publication. The key differences are the considerably
smaller $J_2/J_1$ ratio and coupling to higher neighbors in \ligo. Finally, we
note that both these materials have a substantial 3D coupling, which
leads to long-range order at finite T. It would be interesting to find
a material with large-$J_2$ that was nearly 2D, thus closer to
exhibiting purely quantum order by disorder.

We thank F. Mila and P. Carretta for discussions.  This work was
supported by the DAAD (individual grant H.R.), the NSF DMR-9802076 and
DMR-9986948, the Deutsche Forschungsgemeinschaft, and by the
Australian Research Council.

\end{document}